\documentclass{elsarticle}

\usepackage{amsmath}
\usepackage{color}
\usepackage{float}
\usepackage{amsmath}
\usepackage{array}
\usepackage{rotating}
\usepackage{algpseudocode}
\usepackage{algorithm}
\usepackage{subfigure}
\usepackage{soul}
\mathchardef\mhyphen="2D

\usepackage{changes}
\definechangesauthor[name={Stefano Baroni}, color=orange]{SB}

\def\QE{\textsc{Quantum ESPRESSO}}
\def\SBR{\xrightarrow{\mathrm{SBR}}}

\begin{document}

\begin{frontmatter}

\title{turboEELS -- A code for the simulation of the electron energy loss and inelastic X-ray scattering spectra using the Liouville-Lanczos approach to time-dependent density-functional perturbation theory} 
\author[ecole-polytechnique,sissa]{Iurii Timrov\fnref{fn1}}
\author[ecole-polytechnique]{Nathalie Vast}
\author[ictp]{Ralph Gebauer}
\author[sissa]{Stefano Baroni\corref{author}}

\cortext[author] {Corresponding author. \textit{e-mail address:} baroni@sissa.it}
\address[ecole-polytechnique]{Laboratoire des Solides Irradi\'{e}s, \'Ecole Polytechnique - CEA - DSM - IRAMIS
  - CNRS UMR 7642,  91128 Palaiseau cedex, France}
\address[sissa]{SISSA -- Scuola Internazionale Superiore di Studi Avanzati, Via
  Bonomea 265, 34136 Trieste, Italy} 
\address[ictp]{ICTP -- The Abdus Salam International Centre for Theoretical
  Physics, Strada Costiera 11, 34151 Trieste, Italy}
\fntext[fn1]{Present address: SISSA -- Scuola Internazionale Superiore di Studi Avanzati, Via
  Bonomea 265, 34136 Trieste, Italy}


\begin{abstract}
We introduce \texttt{turboEELS}, an implementation of the Liouville-Lanczos approach to linearized time-dependent density-functional theory, designed to simulate electron energy loss and inelastic X-ray scattering spectra in periodic solids. \texttt{turboEELS} is open-source software distributed under the terms of the GPL as a component of \QE. As with other components, \texttt{turboEELS} is optimized to run on a variety of different platforms, from laptops to massively parallel architectures, using native mathematical libraries (LAPACK and FFTW) and a hierarchy of custom parallelization layers built on top of MPI. 
\end{abstract}

\end{frontmatter}
{\bf PROGRAM SUMMARY}

\begin{small}
  \noindent
  {\em Program Title:} \texttt{turboEELS} \\
  {\em Catalogue identifier:}                                   \\
  {\em Program summary URL:}                                    \\
  {\em Program obtainable from:} CPC Program Library, Queen's University, Belfast, N. Ireland  \\
  {\em Licensing provisions:} GNU General Public License V 2.0  \\
  {\em No. of lines in distributed program, including test data, etc.:} 15\,000 + libraries \\
  {\em No. of bytes in distributed program, including test data, etc.:} 6\,231\,803     \\
  {\em Distribution format:} tar.gz     \\
  {\em Programming language:} Fortran 95                         \\
  {\em Computer:} Any computer architecture                      \\
  {\em Operating system:} GNU/Linux, AIX, IRIX, Mac OS X, and other
  UNIX-like OS's            \\
  {\em Classification:} 7.2 \\
  {\em External routines:} \texttt{turboEELS} is a tightly integrated component of the
  \QE\, distribution and requires the standard libraries linked by it: BLAS, LAPACK, FFTW, MPI.\\
  {\em Nature of problem:} Calculation of the electron energy loss and inelastic X-ray scattering spectra of periodic solids.  
  \\
  {\em Solution method:} The charge-density susceptibility of a periodic system is expressed in terms of the resolvent of its Liouvillian superoperator within time-dependent density functional perturbation theory. It is calculated using non-Hermitian or pseudo-Hermitian variants of the Lanczos recursion scheme, whose implementation does not require the calculation of any virtual states. Pseudopotentials (both norm-conserving and ultrasoft) are used in conjunction with plane-wave basis sets and periodic boundary conditions. Relativistic effects (spin-orbit coupling) can be included in calculations.
  \\
  {\em Restrictions:} Linear-response regime. Adiabatic exchange-correlation kernels only. No hybrid functionals. Collinear spin-polarized formalism is not supported, only non-collinear spin-polarized case can be used. Spin-orbit coupling cannot be used with ultrasoft pseudopotentials. No magnetism. No Hubbard U formalism. No PAW pseudopotentials. \\
  {\em Unusual features:} No virtual orbitals are used, nor even calculated. A single Lanczos recursion gives access to the whole spectrum at fixed transferred momentum. \\
  {\em Additional comments:} The distribution file of this program can be downloaded from the \QE\, website: http://www.quantum-espresso.org/, and the development version of this program can be downloaded via SVN from the QE-forge website: http://qe-forge.org/gf/project/q-e/ . \\
  {\em Running time:} From a few minutes for elemental bulk systems with a few atoms in the primitive unit cell on serial machines up to many hours on multiple processors for complex systems ({\it e.g.}, surfaces with high Miller indices) with dozens or hundreds of atoms. \\
\end{small}

\section{Introduction}
\label{sec:intro}

Electron energy loss (EEL) and inelastic X-ray scattering (IXS) spectroscopies are two experimental techniques
 that probe collective (\emph{plasmon-like}) and single-particle charge excitations
\cite{Egerton:1996, Schulke:2007} in extended systems. Nowadays, EEL experiments with transmission electron microscopes allow one to reach an extremely high spatial resolution of $\sim 0.1$~nm and an energy resolution of typically 1~eV, which can be reduced down to $\sim 0.1$~eV if an electron-beam monochromator is used \cite{Mazzucco:2011, Rivacoba:2000, Garcia:2008}. The EEL and IXS cross sections are directly related to the imaginary part of the inverse of the dielectric function, and thus many electronic properties of the probed materials can be determined. On the theoretical side, time-dependent (TD) density functional theory (DFT) \cite{Runge:1984,Gross:1996, Marques:2012} is the state-of-the-art method that allows one to gain access to collective and single-particle excitations, 
resulting in an optimal compromise between computational cost and accuracy \cite{Onida:2002}.

In the ``conventional'' TDDFT approach to EEL and IXS spectroscopies, the information about plasmon and single-particle excitations is obtained from the poles of the charge-density susceptibility, by first computing the independent-electron susceptibility and solving then a Dyson-like screening equation for the full response function. However, such an approach has several drawbacks: i) the calculation of the independent-electron susceptibility requires the knowledge of numerous empty states, which
are hardly available for systems larger than a few handfuls of independent atoms;
ii) the solution of the Dyson-like screening equation requires multiple time-consuming inversions and multiplications of large matrices; and iii) the first two steps must be repeated for every value of frequency one is interested in, making it difficult to study extended frequency ranges.

Recently a new technique was proposed to overcome these drawbacks,
based on linearized TDDFT (Time-Dependent Density Functional Perturbation Theory, TDDFpT) \cite{Timrov:2013b, Timrov:2013,Walker:2007,Rocca:2008}, 
and inspired by the Liouville-Lanczos (LL) approach to optical spectroscopies in molecular systems, which is in turn the extension to the dynamical regime of time-independent density-functional perturbation theory (DFPT) \cite{Baroni:1987, Baroni:2001}. The distinctive feature of the new approach is that it allows for the full spectrum of a periodic solid to be computed over a broad frequency range,
with a computational effort that is only a few times larger than that needed by a single ground-state DFT or static DFPT calculation at any given finite transferred momentum.

In this paper we introduce a computer code, named \texttt{turboEELS}, which implements the LL approach to EEL and IXS spectroscopies within TDDFpT. \texttt{turboEELS} has a similar structure as the \texttt{turboTDDFT} code
used to compute absorption spectra in molecular systems \cite{Malcioglu:2011, Ge:2014}, and these two codes 
share indeed a large number of routines (in addition, \texttt{turboEELS} also leverages a certain number of solid-state specific routines  from the \texttt{ph.x} component of \QE).
\texttt{turboEELS} is distributed under the terms of the GPL license \cite{GPL}, as a component of the \QE\, suite of open-source codes based on plane-wave basis sets, pseudopotentials, and using periodic boundary conditions \cite{Giannozzi:2009}. 
 
This paper is organized as follows. In Sec.~\ref{sec:theory} we provide a theoretical background for the LL approach to EEL and IXS spectroscopies.
In Sec.~\ref{sec:Description_of_software_components} we describe the components of \texttt{turboEELS} which is a part of the \QE \, distribution. In Sec.~\ref{sec:installation_and_parallelization} we provide the instructions for installing \texttt{turboEELS} on UNIX systems and discuss various levels of parallelization implemented in it.
In Sec.~\ref{sec:benchmarking} we give an example of the usage of \texttt{turboEELS} for the calculation of the EEL spectra of diamond. Finally, Sec.~\ref{sec:Conclusions} contains conclusions and perspectives for future work. 

\section{Theory}
\label{sec:theory}

\subsection{Statement of the problem}
\label{sec:Statement_of_the_problem}

Both EEL and IXS spectroscopies probe the electronic structure of the materials~\cite{Huotari:2012}. In an EEL experiment,
an electron with wave vector $\mathbf{k}_i$ and energy $E_i=\frac{\hbar^2k_i^2}{2m}$ undergoes an inelastic scattering
with the electrons and ions of the target sample, which modifies the electronic wave vector and energy to
$\mathbf{k}_f  = \mathbf{k}_i-\mathbf{Q}$ and $E_f = E_i-\hbar d\omega$, where $\hbar\mathbf{Q}$  
and $\hbar d\omega$ are the momentum and energy transferred to the sample, respectively. Analogously, in an IXS experiment an X-ray photon of energy $E_i=\hbar \omega_i$ and polarization $\mathbf{e}_i$ is scattered to a final state of energy $E_f = \hbar\omega_f$  and polarization $\mathbf{e}_f$. The corresponding double-differential cross section
reads \cite{Egerton:1996,Schulke:2007}:
\begin{equation}
  \frac{d^2\sigma}{d\Omega d\omega} = A \, S(\mathbf{Q},\omega) , 
  \label{eq:cross_section}
\end{equation}
where
$S(\mathbf{Q},\omega)$ is the {\it dynamic structure factor} (per unit volume) of the target sample and
\begin{align}
 A_\mathrm{EEL} &= \left( \frac{4 \pi e^2}{Q^2} \right)^2 \frac{m^2}{4 \pi^2 \hbar^4} \frac{k_f}{k_i} \,,  \\
 A_\mathrm{IXS} &= \left( \frac{e^2}{m c^2} \right)^2 (\mathbf{e}_i \cdot
  \mathbf{e}_f)^2 \, \frac{\omega_f}{\omega_i}  \,,
  \label{eq:cross_section_IXS}
\end{align}
is the {\it probe factor}, whose form differs for the two spectroscopies.
In the above expressions, $-e$ and $m$ are the electron charge and mass, respectively, and $c$ is the speed of light.

According to the fluctuation-dissipation theorem the dynamic structure factor, which describes charge-density fluctuations in the system, is related to the imaginary part of the charge-density susceptibility,  $\chi(\mathbf{Q},\mathbf{Q}'; \omega)$, which describes energy dissipation, by the relation \cite{Pines:1966}:
\begin{equation}
  S(\mathbf{Q},\omega) = - \frac{\hbar}{\pi} \,
  \mathrm{Im} \, \chi(\mathbf{Q},\mathbf{Q}; \omega) . 
  \label{eq:fluct_dissip_theorem}
\end{equation}
The poles of $\chi(\mathbf{Q},\mathbf{Q}; \omega)$ correspond to plasmon frequencies and single-particle excitations.

In periodic solids the transferred momentum, $\mathbf{Q}$, can be expressed as the sum of an {\it irreducible component} lying inside of the first Brillouin zone (BZ), $\mathbf{q}$, and a reciprocal lattice vector, $\mathbf{G}$: $\mathbf{Q}=\mathbf{q}+\mathbf{G}$. 
The EEL/IXS cross section is often expressed in terms of the inverse dielectric matrix, $\epsilon^{-1}_{\mathbf{G}\mathbf{G}'}(\mathbf{q};\omega) \equiv \epsilon^{-1}(\mathbf{q}+\mathbf{G},\mathbf{q}+\mathbf{G}';\omega)$, defined as
\cite{Martin:2004, Car:1981}:
\begin{equation}
  \epsilon^{-1}(\mathbf{q}+\mathbf{G},\mathbf{q}+\mathbf{G}';\omega)  = 
  \delta_{\mathbf{G}\mathbf{G}'} + \frac{4\pi e^2}{|\mathbf{q}+\mathbf{G}|^2} \, 
  \chi(\mathbf{q}+\mathbf{G},\mathbf{q}+\mathbf{G}';\omega) .
  \label{eq:def_inv_microscopic_diel_tensor}
\end{equation}
The diagonal elements of the inverse dielectric matrix define the inverse dielectric function, $\epsilon^{-1}(\mathbf{Q},\omega) \equiv \epsilon^{-1}(\mathbf{q}+\mathbf{G},\mathbf{q}+\mathbf{G};\omega)$, whose imaginary part 
is the negative of the so-called {\it loss function}, and which satisfies the Thomas-Reiche-Kuhn (or $f$-sum) rule \cite{Mahan:1975}:
\begin{equation}
   \int\limits_0^\infty \mathrm{Im}[\epsilon^{-1}(\mathbf{Q},\omega)] \, 
   \omega \, d\omega = -\frac{\pi}{2} \omega_p^2 ,
   \label{eq:f-sum_definition}
\end{equation}
where $\omega_p = \sqrt{4\pi e^2 n/m}$ is the Drude plasma frequency, where $n$ is the average electron density. The dielectric function is {\it defined} as the inverse of the inverse dielectric function, $\epsilon(\mathbf{Q},\omega) \equiv 1/\epsilon^{-1}(\mathbf{Q},\omega)$ (Note: this is not a matrix inversion). The information about the real and imaginary parts of $\epsilon(\mathbf{Q},\omega)$ can be very useful for an analysis of the EEL/IXS spectra ({\it i.e.} the loss function), because the loss function, $-\mathrm{Im}[\epsilon^{-1}(\mathbf{Q},\omega)] = \mathrm{Im}[\epsilon(\mathbf{Q},\omega)]/ \left( \mathrm{Re}[\epsilon(\mathbf{Q},\omega)]^2 + \mathrm{Im}[\epsilon(\mathbf{Q},\omega)]^2 \right)$, has peaks when $\mathrm{Re}[\epsilon(\mathbf{Q},\omega)]=0$ and $\mathrm{Im}[\epsilon(\mathbf{Q},\omega)]$ is small. For small $\mathbf{Q}$ the plasmon occurs when $\mathrm{Re}[\epsilon(\mathbf{Q},\omega)]=0$ and it has a positive slope (changes the sign from negative to positive).

\subsection{Quantum Liouville equation for the charge-density susceptibility} 
\label{sec:susceptibility_def}

As we have seen in the previous section, the EEL/IXS cross section is essentially determined by the diagonal of the charge-density susceptibility, $\chi(\mathbf{Q},\mathbf{Q};\omega)$. The latter quantity is the response of the Fourier transform of the charge-density operator, $\hat{n}_{\mathbf{Q}} \rightarrow \mathrm{e}^{i\mathbf{Q}\cdot \mathbf{r}}$, to a monochromatic perturbation of same wave vector, $\hat{v}'_{\mathrm{ext},\mathbf{q}}=\hat{n}_{\mathbf{Q}}$. Here and in the following quantum-mechanical operators and superoperators are denoted by a hat ``\^{}'', whereas their coordinate representation does not have any hat. Here $\mathbf{q}$ denotes the irreducible component of $\mathbf{Q}$ and the notation has been chosen so as to emphasize that the response to a monochromatic perturbation is in general not monochromatic, while conserving the irreducible component of its wave vector. 
In particular, the response of the density matrix to a monochromatic perturbation, $\hat{v}'_{\mathrm{ext},\mathbf{q}}=\hat{n}_{\mathbf{Q}}$, can be labeled by the irreducible component of its wave vector, $\hat{\rho}'_{\mathbf{q}}$, and so can the response of the Hartree and exchange-correlation (HXC) potential, $\hat{v}'_{\mathrm{HXC},\mathbf{q}}[\hat{\rho}'_\mathbf{q}]$ 
(\emph{i.e.} they result to be the product of a lattice-periodic function times a plane wave with wave vector $\mathbf{q}$, $\mathrm{e}^{i\mathbf{q}\cdot\mathbf{r}}$). In terms of $\hat{\rho}'_\mathbf{q}$, the relevant matrix element of the charge-density susceptibility reads:
\begin{equation}
\chi(\mathbf{Q},\mathbf{Q};\omega) = \mathrm{Tr} \left( \hat{n}_\mathbf{Q}^\dagger \, \hat{\rho}'_{\mathbf{q}} \right) .
\end{equation}
As it was shown in Refs.~\cite{Walker:2007,Rocca:2008,Baroni:2012b}, the response density matrix 
satisfies a superoperator linear equation -- {\it the quantum Liouville equation} -- that can be symbolically cast as:
\begin{equation}
  (\omega - \hat{\mathcal{L}}) \cdot \hat{\rho}^\prime_\mathbf{q}(\omega) =
  [\hat{v}'_{\mathrm{ext},\mathbf{q}}, \hat{\rho}^\circ ] ,
  \label{eq:Liouville_eq_FT_1_general}
\end{equation}
where the action of the Liouvillian superoperator $\hat{\mathcal{L}}$ onto $\hat{\rho}^\prime_\mathbf{q}$ is defined as:
\begin{equation}
  \hat{\mathcal{L}} \cdot \hat{\rho}^\prime_\mathbf{q} \equiv [ \hat{H}^\circ, \hat{\rho}^\prime_\mathbf{q} ] +
  [ \hat{v}'_{\mathrm{HXC},\mathbf{q}}[\hat{\rho}'_\mathbf{q}], \hat{\rho}^\circ ] ,
  \label{eq:Liouvillian_def}
\end{equation}
and Hartree atomic units ($e=m=\hbar=1$) are used henceforth. Here $[\cdot,\cdot]$ indicates a quantum commutator, $\hat{H}^\circ$ and $\hat{\rho}^\circ$ are the unperturbed Kohn-Sham (KS) Hamiltonian and density matrix, respectively. 
According to the above considerations, the relevant matrix element of the charge-density susceptibility can be represented as a suitable off-diagonal matrix element of the resolvent of the Liouvillian superoperator as:
\begin{equation}
  \chi(\mathbf{Q},\mathbf{Q}; \omega) = 
   \left( \hat{n}_\mathbf{Q} , (\omega - \hat{\mathcal{L}})^{-1} \cdot [\hat{n}_{\mathbf{Q}}, \hat{\rho}^\circ ] \right) .
  \label{eq:susceptibility_def_2}
\end{equation}
The essential ingredients for evaluating Eq.~\eqref{eq:susceptibility_def_2} are thus an explicit representation for the response density matrix and an efficient way of evaluating off-diagonal matrix elements of the resolvent of linear (super-) operators. The subsections \ref{sec:SBR} and \ref{sec:Lanczos_algorithm} are devoted to these two tasks, respectively. The problem of computing the charge-density susceptibility \eqref{eq:susceptibility_def_2} is mapped onto the solution of a set of coupled linearized KS equations in the self-consistent way, as will be shown in Sec.~\ref{sec:Bloch_functs}.

\subsection{Bloch functions and coordinate representation of (response) operators}
\label{sec:Bloch_functs}

In periodic solids the unperturbed KS orbitals, which satisfy the ground-state KS equations, have a Bloch form: 
$\varphi^\circ_{n,\mathbf{k}}(\mathbf{r}) = \mathrm{e}^{i\mathbf{k}\cdot\mathbf{r}} \,\mathrm{u}^\circ_{n,\mathbf{k}}(\mathbf{r})$, 
where $n$ is a band index, $\mathbf{k}$ is a point in BZ, and $\mathrm{u}^\circ_{n,\mathbf{k}}(\mathbf{r})$ is a 
lattice-periodic function. The response KS 
orbitals corresponding to each monochromatic $\mathbf{q}$ component of the perturbing potential, 
$\hat v'_{\mathrm{ext},\mathbf{q}}$, read:
$\tilde\varphi'_{n,\mathbf{k}+\mathbf{q}}(\mathbf{r},\omega) = \mathrm{e}^{i(\mathbf{k}+\mathbf{q}) 
\cdot \mathbf{r}} \tilde{\mathrm{u}}'_{n,\mathbf{k+q}}(\mathbf{r},\omega)$,
where $\tilde{\mathrm{u}}'_{n,\mathbf{k+q}}(\mathbf{r},\omega)$ is the lattice-periodic response orbital. The response and perturbing potentials in the coordinate representation read as $\tilde{v}'_\mathbf{q}(\mathbf{r},\omega) = \mathrm{e}^{i\mathbf{q}\cdot\mathbf{r}} \, \tilde{\mathrm{v}}'_\mathbf{q}(\mathbf{r},\omega)$, where $\tilde{\mathrm{v}}'_\mathbf{q}(\mathbf{r},\omega)$ is the lattice-periodic part of the potential.  
Here and below, Roman letters in formulas indicate lattice-periodic parts of Bloch functions, and tildes ``$\tilde{\phantom{a}}$'' indicate Fourier transforms of functions in the frequency domain. 

It can be demonstrated that the lattice-periodic response orbitals, $\tilde{\mathrm{u}}'_{n,\mathbf{k+q}}(\mathbf{r},\omega)$, satisfy the linearized and Fourier transformed KS equations \cite{Timrov:2013b}:
\begin{eqnarray}
  (\hat{\mathrm{H}}^\circ_\mathbf{k+q} - \varepsilon^\circ_{n,\mathbf{k}} - \omega) \,
  \tilde{\mathrm{u}}'_{n,\mathbf{k+q}}(\mathbf{r},\omega) & + & 
  \hat{\mathrm{P}}_c^\mathbf{k+q} \, \tilde{\mathrm{v}}'_{\mathrm{HXC},\mathbf{q}}(\mathbf{r},\omega)
  \, \mathrm{u}_{n,\mathbf{k}}^\circ(\mathbf{r}) \nonumber  \\ 
  & = & - \hat{\mathrm{P}}_c^\mathbf{k+q} \, \tilde{\mathrm{v}}'_{\mathrm{ext},\mathbf{q}}(\mathbf{r},\omega)
  \, \mathrm{u}_{n,\mathbf{k}}^\circ(\mathbf{r}) ,
  \label{eq:lin-resp_1}
\end{eqnarray}
where $\varepsilon^\circ_{n,\mathbf{k}}$ are the ground-state KS energies, $\tilde{\mathrm{v}}'_{\mathrm{ext},\mathbf{q}}(\mathbf{r},\omega) = \mathrm{e}^{i\mathbf{G}\cdot\mathbf{r}}$ is the lattice-periodic part of the external perturbing potential $\tilde{v}'_{\mathrm{ext},\mathbf{q}}(\mathbf{r},\omega)$, and $\tilde{\mathrm{v}}'_{\mathrm{HXC},\mathbf{q}}$ is the lattice-periodic $\mathbf{q}$ component of the HXC response potential, which reads:
\begin{equation}
  \tilde{\mathrm{v}}'_{\mathrm{HXC},\mathbf{q}}(\mathbf{r},\omega) =
  \int\kappa(\mathbf{r},\mathbf{r}') \, 
  \tilde{\mathrm{n}}'_\mathbf{q}(\mathbf{r}',\omega) \, \mathrm{e}^{-i\mathbf{q}\cdot(\mathbf{r} - \mathbf{r}')} d\mathbf{r}',
  \label{eq:V_HXC_resp_1}
\end{equation}
where $\kappa(\mathbf{r},\mathbf{r}') = 1/|\mathbf{r-r'}| + \kappa_\mathrm{XC}(\mathbf{r},\mathbf{r}')$, $\kappa_\mathrm{XC}(\mathbf{r},\mathbf{r}')$ is the XC kernel \cite{Gross:1985}, which in the adiabatic DFT approximation adopted in this paper is independent of frequency, and $\tilde{\mathrm{n}}^\prime_\mathbf{q}(\mathbf{r}, \omega)$ is the lattice-periodic $\mathbf{q}$ component of the response charge-density.
In practice, the response XC potential is computed in the real space, whereas the response Hartree potential is computed in the reciprocal space as: 
\begin{equation}
  \tilde{\mathrm{v}}'_{\mathrm{H},\mathbf{q}}(\mathbf{r},\omega) = 4\pi
  \sum\limits_\mathbf{\mathbf{G}}
  \frac{\tilde{\mathrm{n}}'_\mathbf{q}(\mathbf{G},\omega)}{|\mathbf{q}+\mathbf{G}|^2}
  \, \mathrm{e}^{i\mathbf{G}\cdot\mathbf{r}} .
  \label{eq:V_HXC_resp_2}
\end{equation}
In Eq.~\eqref{eq:lin-resp_1}, as well as in the rest of this paper, quantum-mechanical operators having a wave vector subscript (such as $\hat{\mathrm{H}}^\circ_\mathbf{k+q}$) or superscript (such as $\hat{\mathrm{P}}_c^\mathbf{k+q}$) operate on lattice-periodic functions, and they are defined in terms of their coordinate representations as:
\begin{align}
  H^\circ(\mathbf{r},\mathbf{r}') &= \sum_{\mathbf{k}} \mathrm{e}^{i
    \mathbf{k}\cdot(\mathbf{r}-\mathbf{r}')}\, \mathrm{H}^\circ_\mathbf{k}
  (\mathbf{r},\mathbf{r}')\,, \\
  P_c(\mathbf{r},\mathbf{r}') &= \sum_{\mathbf{k}} \mathrm{e}^{i
    \mathbf{k}\cdot(\mathbf{r}-\mathbf{r}')}
  \mathrm{P}_c^{\mathbf{k}}(\mathbf{r},\mathbf{r}'),
  \label{eq:Projector_on_empty_states_2}
\end{align}
where the projector onto the conduction manifold, $\mathrm{P}_c^{\mathbf{k}}(\mathbf{r},\mathbf{r}')$, can be expressed in terms of the lattice-periodic parts of the unperturbed Bloch functions as:
\begin{equation}
  \mathrm{P}_c^{\mathbf{k}}(\mathbf{r,r'})  = \delta (\mathbf{r-r'}) - \sum_n^{occ}
  \mathrm{u}^\circ_{n,\mathbf{k}}(\mathbf{r}) \,
  \mathrm{u}^{\circ\,*}_{n,\mathbf{k}}(\mathbf{r}') , 
\end{equation}
where the sum extends over all the occupied bands. Thus, in Eq.~\eqref{eq:lin-resp_1} no empty states are needed to determine the response orbitals, much in the same way like in static density functional perturbation theory (DFPT) \cite{Baroni:1987, Baroni:2001}.

The monochromatic $\mathbf{q}$ component of the response density matrix in the coordinate representation reads:
\begin{eqnarray}
  \tilde{\rho}'_\mathbf{q}(\mathbf{r},\mathbf{r}';\omega) 
  & = & 2 \sum_{n,\mathbf{k}} \Bigl(
  \tilde{\varphi}^\prime_{n,\mathbf{k+q}}(\mathbf{r},\omega) \,
  \varphi^{\circ\,*}_{n,\mathbf{k}}(\mathbf{r}') +  
  \varphi^{\circ\,*}_{n,\mathbf{k}}(\mathbf{r}) \,
  \tilde{\varphi}^{\prime\,*}_{n,\mathbf{-k-q}}(\mathbf{r}',-\omega) \Bigr) \nonumber \\
  & = & 2 \sum_{n,\mathbf{k}} \Bigl(
  \mathrm{e}^{i\mathbf{q}\cdot\mathbf{r}} \,
  \tilde{\mathrm{u}}^\prime_{n,\mathbf{k+q}}(\mathbf{r},\omega) \, \mathrm{u}^{\circ\,*}_{n,\mathbf{k}}(\mathbf{r}') \,
  \mathrm{e}^{i\mathbf{k}\cdot(\mathbf{r}-\mathbf{r}')} \nonumber \\
  &  & \hspace{0.6 cm} + \, \mathrm{e}^{i\mathbf{q}\cdot\mathbf{r}'} \, 
  \mathrm{u}^{\circ\,*}_{n,\mathbf{k}}(\mathbf{r}) \,
  \tilde{\mathrm{u}}^{\prime\,*}_{n,\mathbf{-k-q}}(\mathbf{r}',-\omega) \,
  \mathrm{e}^{- i\mathbf{k}\cdot(\mathbf{r}-\mathbf{r}')} \Bigr) ,
  \label{eq:rho_prime-periodic}
\end{eqnarray}
where the factor of 2 is due to spin degeneracy. Hence, the response charge-density, which is a diagonal of the response density matrix, $\tilde{n}'_\mathbf{q}(\mathbf{r},\omega) = \tilde{\rho}'_\mathbf{q}(\mathbf{r},\mathbf{r};\omega)$, reads:
\begin{equation}
\tilde{n}'_\mathbf{q}(\mathbf{r},\omega) = \mathrm{e}^{i\mathbf{q}\cdot\mathbf{r}} \, \tilde{\mathrm{n}}'_\mathbf{q}(\mathbf{r},\omega) ,
\end{equation}
where $\tilde{\mathrm{n}}'_\mathbf{q}(\mathbf{r},\omega)$ is the lattice-periodic charge-density, which reads:
\begin{equation}
  \tilde{\mathrm{n}}'_\mathbf{q}(\mathbf{r},\omega) =
  2 \sum_{n,\mathbf{k}} \mathrm{u}^{\circ\,*}_{n,\mathbf{k}}(\mathbf{r}) \Bigl(
  \tilde{\mathrm{u}}^\prime_{n,\mathbf{k+q}}(\mathbf{r},\omega) +
  \tilde{\mathrm{u}}^{\prime\,*}_{n,\mathbf{-k-q}}(\mathbf{r},-\omega) \Bigr) .
  \label{eq:rho_prime-periodic_2}
\end{equation}
The antiresonant contribution to the response density matrix \eqref{eq:rho_prime-periodic} satisfies the equation:
\begin{eqnarray}
  (\hat{\mathrm{H}}^\circ_\mathbf{k+q} - \varepsilon^\circ_{n,\mathbf{k}} + \omega) \,
  \tilde{\mathrm{u}}^{\prime\,*}_{n,\mathbf{-k-q}}(\mathbf{r},-\omega) & + & 
  \hat{\mathrm{P}}_c^\mathbf{k+q} \, \tilde{\mathrm{v}}'_{\mathrm{HXC},\mathbf{q}}(\mathbf{r},\omega)
  \, \mathrm{u}_{n,\mathbf{k}}^\circ(\mathbf{r}) \nonumber  \\ 
  & = & - \hat{\mathrm{P}}_c^\mathbf{k+q} \, \tilde{\mathrm{v}}'_{\mathrm{ext},\mathbf{q}}(\mathbf{r},\omega)
  \, \mathrm{u}_{n,\mathbf{k}}^\circ(\mathbf{r}) ,
  \label{eq:lin-resp_2}
\end{eqnarray}
which can be obtained from Eq.~\eqref{eq:lin-resp_1} by complex conjugation and simple manipulations deriving from time-reversal invariance of the unperturbed system [$\mathrm{u}^\circ_{n,\mathbf{k}}(\mathbf{r}) = \mathrm{u}^{\circ\,*}_{n,\mathbf{-k}}(\mathbf{r})$] and the reality of the perturbing and HXC response potentials ($ \tilde{\mathrm{v}}'_{\mathrm{ext},\mathbf{q}}(\mathbf{r},\omega) = \tilde{\mathrm{v}}^{\prime \, *}_{\mathrm{ext},\mathbf{-q}}(\mathbf{r},-\omega)$ and $ \tilde{\mathrm{v}}'_{\mathrm{HXC},\mathbf{q}}(\mathbf{r},\omega) = \tilde{\mathrm{v}}^{\prime \, *}_{\mathrm{HXC},\mathbf{-q}}(\mathbf{r},-\omega)$, respectively). The set of coupled KS equations~\eqref{eq:lin-resp_1} and \eqref{eq:lin-resp_2} represent the quantum Liouville equation~\eqref{eq:Liouville_eq_FT_1_general}.

\subsection{Standard batch representation of operators}
\label{sec:SBR}

Equations~\eqref{eq:lin-resp_1}, \eqref{eq:rho_prime-periodic} and \eqref{eq:lin-resp_2} show that the response of the charge-density of a periodic solid to a perturbation of wave vector $\mathbf{Q}$ is uniquely determined by the two sets of response orbitals $\{ \tilde{\mathrm{u}}'_{n,\mathbf{k+q}} (\mathbf{r}, \omega) \}$ and $\{ \tilde{\mathrm{u}}^{\prime\,*}_{n,\mathbf{-k-q}} (\mathbf{r}, - \omega)\}$. Note that $n$ and $\mathbf{k}$ are running indices, whereas $\mathbf{q}$ is fixed. It is convenient to consider a linear combination of these functions, defined as 
\begin{align}
  \mathrm{q}_{n,\mathbf{k+q}}(\mathbf{r}) &= \frac{1}{2} \, \bigl (
  \tilde{\mathrm{u}}'_{n,\mathbf{k+q}}(\mathbf{r},\omega) +
  \tilde{\mathrm{u}}^{\prime\,*}_{n,\mathbf{-k-q}}(\mathbf{r},-\omega) \bigr ) , 
  \\
  \mathrm{p}_{n,\mathbf{k+q}}(\mathbf{r}) &= \frac{1}{2} \, \bigl (
  \tilde{\mathrm{u}}'_{n,\mathbf{k+q}}(\mathbf{r},\omega) -
  \tilde{\mathrm{u}}^{\prime\,*}_{n,\mathbf{-k-q}}(\mathbf{r},-\omega) \bigl ) . 
\end{align}
The two sets of response orbitals $\mathrm{q}_\mathbf{q}=\{\mathrm{q}_{n,\mathbf{k+q}}\} $ and $\mathrm{p}_\mathbf{q}=\{\mathrm{p}_{n,\mathbf{k+q}}\}$ are called, respectively, the upper and lower components of the {\it standard batch representation} (SBR) \cite{Rocca:2008, Baroni:2012b} of the response density matrix supervector: $\hat\rho'_\mathbf{q} \SBR \{ \mathrm{q}_\mathbf{q}, \mathrm{p}_\mathbf{q} \}$ \cite{Malcioglu:2011}. 
In the SBR the set of Eqs.~\eqref{eq:lin-resp_1} and \eqref{eq:lin-resp_2} can be cast into the compact form:
\begin{equation}
  \left( \omega \hat{I} - \hat{\mathcal{L}}_{\mathbf{q}} \right)
  \left(
    \begin{array}{c}
      \mathrm{q}_\mathbf{q} \\
      \mathrm{p}_\mathbf{q}
    \end{array}
  \right) =
  \left(
    \begin{array}{c}
      0 \\
      \mathrm{y}_\mathbf{q}
    \end{array}
  \right) ,
  \label{eq:Liouvillian_eq_SBR_2}
\end{equation}
where $\hat{I}$ is the unit matrix, and the Liouvillian $\hat{\mathcal{L}}_{\mathbf{q}}$ has the block form \cite{Rocca:2008, Baroni:2012b}:
\begin{equation}
  \hat{\mathcal{L}}_{\mathbf{q}} =
  \left(
    \begin{array}{cc}
      0 & \hat{\mathcal{D}}_\mathbf{q} \\ 
      \hat{\mathcal{D}}_\mathbf{q} + \hat{\mathcal{K}}_\mathbf{q} & 0
    \end{array}
  \right) ,
  \label{eq:Liouvillian_SBR}
\end{equation}
where $\hat{\mathcal{D}}_\mathbf{q}$ and $\hat{\mathcal{K}}_\mathbf{q}$ superoperators are defined by their action on batches,
\begin{align}
  \hat{\mathcal{D}}_\mathbf{q} \mathrm{q}_\mathbf{q} & = \left \{
    (\hat{\mathrm{H}}^\circ_\mathbf{k+q} - \varepsilon^\circ_{n,\mathbf{k}}) \,
    \mathrm{q}_{n,\mathbf{k+q}}(\mathbf{r}) \right \} , 
    \label{eq:D_super-operator_2} \\
    \hat{\mathcal{K}}_\mathbf{q} \mathrm{q}_\mathbf{q} & 
    = \left \{ 
    \hat{\mathrm{P}}_c^\mathbf{k+q} \,
    \tilde{\mathrm{v}}'_{\mathrm{HXC},\mathbf{q}}(\mathbf{r}) \,
    \mathrm{u}^\circ_{n,\mathbf{k}}(\mathbf{r})  \right \},
    \label{eq:K_super-operator_2}
\end{align}
and $\tilde{\mathrm{v}}'_{\mathrm{HXC},\mathbf{q}}$ is given by Eq.~\eqref{eq:V_HXC_resp_1}, and the lattice-periodic response charge-density reads (see Eq.~\eqref{eq:rho_prime-periodic_2}):
\begin{equation}
  \tilde{\mathrm{n}}'_\mathbf{q}(\mathbf{r}) = 4 \sum_{n,\mathbf{k}}
  \mathrm{u}^{\circ\,*}_{n,\mathbf{k}}(\mathbf{r}) \, \mathrm{q}_{n,\mathbf{k+q}}(\mathbf{r}) ,
  \label{eq:SBR_charge-density_2}
\end{equation}
and
\begin{equation}
  \mathrm{y}_\mathbf{q} = \{ \hat{\mathrm{P}}_c^\mathbf{k+q} \, \tilde{\mathrm{v}}'_{\mathrm{ext},\mathbf{q}}(\mathbf{r}) \, \mathrm{u}_{n,\mathbf{k}}^\circ(\mathbf{r}) \} .
  \label{eq:RHS_of_Liouville_eq}
\end{equation}
According to the above equations, operating with the Liouvillian on a test supervector essentially requires a calculation of the response HXC potential, its application to each valence KS orbital, as well as the operation of the unperturbed Hamiltonian onto twice the number of valence KS states. 

The SBR of the charge-density operator reads: $\hat{n}_\mathbf{Q} \SBR \bigl\{ \{ \hat{\mathrm{P}}_c^\mathbf{k+q} \, \mathrm{e}^{i\mathbf{G\cdot r}} \, \mathrm{u}^\circ_{n,\mathbf{k}} \}, 0 \bigr \}$, and hence 
the SBR of the charge-density susceptibility, Eq.~\eqref{eq:susceptibility_def_2}, reads:
\begin{equation}
  \chi(\mathbf{Q,Q};\omega) = \Bigl ( \bigl\{ \{ 
  \hat{\mathrm{P}}_c^\mathbf{k+q} \, \mathrm{e}^{i\mathbf{G\cdot r}} \, \mathrm{u}^\circ_{n,\mathbf{k}}
  \}, 0 \bigr \}, \bigl \{\mathrm{q}_\mathbf{q},\mathrm{p}_\mathbf{q} \bigr \} \Bigl ),
  \label{eq:susceptibility_Q}
\end{equation}
where $ \bigl \{\mathrm{q}_\mathbf{q},\mathrm{p}_\mathbf{q} \bigr \} $ is the solution of the quantum Liouville equation in the SBR~\eqref{eq:Liouvillian_eq_SBR_2}. In practice, the charge-density susceptibility \eqref{eq:susceptibility_Q} is computed using the Lanczos recursion algorithm, which is briefly discussed in the next subsection.


\subsection{Lanczos recursion algorithm}
\label{sec:Lanczos_algorithm}

From Eq.~\eqref{eq:susceptibility_def_2} it is seen that in order to compute the charge-density susceptibility, $\chi(\mathbf{Q}, \mathbf{Q}; \omega)$, one needs to evaluate the off-diagonal matrix element of the resolvent of the Liouvillian, $(\omega - \hat{\mathcal{L}}_\mathbf{q})^{-1}$. A straightforward inversion of such a matrix for big systems in plane-wave framework is a formidable task. However, there exists an efficient recursive algorithm, so-called {\it Lanczos algorithm}, which does not rely on the inversion of the matrices, but a recursive evaluation of an off-diagonal matrix element as in Eq.~\eqref{eq:susceptibility_def_2} \cite{Saad:2003}. We will briefly review two flavors of the Lanczos algorithm which can be used to compute the charge-density susceptibility, namely, {\it non-Hermitian Lanczos biorthogonalization algorithm} \cite{Rocca:2008, Malcioglu:2011, Baroni:2012b}, and {\it pseudo-Hermitian Lanczos algorithm} \cite{Ge:2014, Gruning:2011, Mostafazadeh:2002}. A detailed description of the algorithms can be found in the corresponding references.

In the non-Hermitian Lanczos biorthogonalization algorithm, by starting from the initial pair of Lanczos vectors $u_1=v_1= \{ 0, \mathrm{y}_\mathbf{q}\}$ (see Eq.~\eqref{eq:RHS_of_Liouville_eq}), two coupled Lanczos chains are generated (because there are two coupled KS equations, resonant and antiresonant \cite{Timrov:2013b}) by recursively applying $\hat{\mathcal{L}}_\mathbf{q}$ and $\hat{\mathcal{L}}_\mathbf{q}^{\dagger}$ to the previous Lanczos chain vectors, $u_i$ and $v_i$ \cite{Timrov:2013b, Rocca:2008}. A pair of biorthogonal basis sets of increasing dimension are thus recursively constructed, $\{ u_i \}$ and $\{ v_i \}$, where $i=\overline{1,M}$, and $M$ being the number of Lanczos iterations, and the Lanczos coefficients, $\beta_i$ and $\gamma_i$, are thus generated form a sparse $M$-dimensional tridiagonal matrix, $^MT_\mathbf{q}$, which is nothing but an oblique projection of the Liouvillian onto such biorthogonal bases: $\left(^MT_{\mathbf{q}}\right)_{ij} = (u_i, \hat{\mathcal{L}}_\mathbf{q} v_j)$.

The pseudo-Hermitian Lanczos algorithm is two times faster than the non-Hermitian Lanczos biorthogonalization algorithm, because the former algorithm requires twice as less operations (action of the Liouvillian, $\hat{\mathcal{L}}_\mathbf{q}$, on the Lanczos vectors) \cite{Ge:2014}. The main idea is to define a metric of the linear space, which redefines the scalar products, and thus allows one to use a generalized Hermitian Lanczos algorithm. As a result of the use of this algorithm one also generates the tridiagonal matrix $^MT_\mathbf{q}$.

After a generation of the tridiagonal matrix, $^MT_\mathbf{q}$, the charge-density susceptibility (see Eqs.~\eqref{eq:susceptibility_def_2} and \eqref{eq:susceptibility_Q}) can be computed as \cite{Timrov:2013b}:
\begin{equation}
  \chi(\mathbf{Q},\mathbf{Q};\omega) \simeq \left(
  ^Mz_\mathbf{q} , ( \omega \, ^MI - \, ^MT_\mathbf{q})^{-1} \cdot \, ^Me_1 \right) , 
  \label{eq:resolvent_Liouvillian_Lanczos_q}
\end{equation}
where $^MI$ is the $M$-dimensional unit matrix, $^Me_1 = \{1,0,\ldots,0\}$ is the $M$-dimensional unit vector, and $^Mz_\mathbf{q} = (z_{1,\mathbf{q}}, z_{2,\mathbf{q}}, \ldots, z_{M,\mathbf{q}})$ is the $M$-dimensional array whose coefficients $z_{j,\mathbf{q}}$ are computed {\it on-the-fly} of the Lanczos recursion and defined as:
\begin{equation}
  z_{j,\mathbf{q}} = \left( \{ \{ \hat{\mathrm{P}}_c^\mathbf{k+q} \,
    \mathrm{e}^{i\mathbf{G}\cdot\mathbf{r}} \, \mathrm{u}_{n,\mathbf{k}}^\circ(\mathbf{r}) \} ,
    0 \} , v_j \right) .
  \label{eq:zeta_coef_q}
\end{equation}
In practice, the right-hand side of Eq.~\eqref{eq:resolvent_Liouvillian_Lanczos_q} is computed by solving, for any given value of frequency, $\omega$, the equation:
\begin{equation}
  \left( \omega \, ^MI - \, ^MT_\mathbf{q} \right) {^Mx}_\mathbf{q} = \, ^Me_1 ,
  \label{eq:post_processing}
\end{equation}
and calculating the scalar product afterwards:
\begin{equation}
  \chi(\mathbf{Q},\mathbf{Q};\omega) = \left ( ^Mz_\mathbf{q} , ^Mx_\mathbf{q} \right ) ,
  \label{eq:resolvent_Liouvillian_Lanczos_2_for_chi}
\end{equation}
which are both extremely inexpensive operations from the computational point of view. This allows one to study a wide frequency range for EEL and IXS. The convergence of the EEL/IXS spectra with respect to the number of Lanczos iterations, $M$, can be sped up by making use of the extrapolation technique for the Lanczos coefficients, which is described in detail in Refs.~\cite{Rocca:2008, Malcioglu:2011}.


\subsection{Extensions of the Liouville-Lanczos approach}

The LL formalism presented above has been generalized in several ways: $\emph{i)}$ metallic systems; $\emph{ii)}$ explicit account of symmetry for $\mathbf{k}$-point sampling; $\emph{iii)}$ relativistic effects (spin-orbit coupling). A detailed discussion about these extensions can be found in Ref.~\cite{Timrov:2013}, and here we discuss briefly only about the basic concepts.

An extension of the LL approach to metals is based on the use of the smearing techniques introduced in the static case for lattice-dynamical calculations \cite{deGironcoli:1995}. In metals, electronic levels crossing the Fermi level do not have integer occupancy (0 or 1), thus smearing techniques must be used to describe the electronic states and allow them to have partial occupancy (between 0 and 1). The LL approach can still have the same form as for non-metallic systems but the following modifications: $\emph{i)}$~the projector on empty states, $\hat{\mathrm{P}}_c^\mathbf{k+q}$, is redefined in a different way, and $\emph{ii)}$~a fractional occupancy of electronic states should be allowed for, by introducing a suitable weight in sums over $\mathbf{k}$ points and band indices. An interested reader can find a detailed discussion in the indicated references.

The CPU time and memory requirements of the LL calculation can be very significantly reduced by employing a symmetry of the system. The space group of the crystal contains such operations as lattice translations and rotations that leave the crystal unchanged. By using these symmetry operations it is possible to reduce the number of $\mathbf{k}$ points and use only those which are in the irreducible wedge of BZ. However, in the case of EEL/IXS the symmetry of the system is reduced by the external monochromatic perturbation with a wave vector~$\mathbf{Q}$. This fact must be taken into account, and thus one can consider $\mathbf{k}$ points inequivalent to each other according to the small group of $\mathbf{Q}$. 

For heavy-atom elements relativistic effects become important and thus must be taken into account for an accurate description of the EEL/IXS spectra. The LL approach has been extended to a treatment of the spin-orbit coupling effect, in a self-consistent way. In this case, instead of the scalar KS orbitals there are two-component spinors, and hence the number of electronic states in the calculation is doubled. Instead of solving non-relativistic linear-response KS equations for scalar KS orbitals in combination with non- or semi-relativistic pseudopotentials, one has to solve Pauli-type linear-response KS equations for spinors and use fully relativistic pseudopotentials. An example of the application of the relativistic LL approach is semimetallic bismuth \cite{Timrov:2013, Timrov:2014b}.


\section{Description of software components}
\label{sec:Description_of_software_components}

The \texttt{turboEELS} code is designed as a module of the \QE\,
distribution \cite{Giannozzi:2009}. It resides in a self contained
directory \texttt{TDDFPT} under the root directory of the
\QE\, tree, which contains also the \texttt{turboTDDFT}
code in its two flavors (Lanczos and Davidson \cite{Ge:2014}) for the
calculation of the absorption spectra. The \texttt{turboEELS} code is
tightly integrated in the 
\texttt{turboTDDFT} code and uses many of its routines. When the
\texttt{turboEELS} code is installed (see
Sec.~\ref{sec:Installation_instructions}), the \texttt{bin/} directory
in the \QE\, root contains links to the executables
\texttt{turbo\_eels.x} (the main program) and
\texttt{turbo\_spectrum.x} (a post-processing
tool). The code \texttt{turbo\_eels.x} performs a Lanczos recursion to obtain
Lanczos coefficients $\beta_i$ and $\gamma_i$, and $z_i$ coefficients (see
Sec.~\ref{sec:Lanczos_algorithm}), and to thus construct a tridiagonal
matrix $^MT_\mathbf{q}$, while \texttt{turbo\_spectrum.x} uses
this matrix to calculate the charge-density susceptibility,
$\chi(\mathbf{Q}, \mathbf{Q}; \omega)$, according to
Eqs.~\eqref{eq:post_processing} and
\eqref{eq:resolvent_Liouvillian_Lanczos_2_for_chi}.

\subsection{Ground-state calculation}   

In order to compute the EEL/IXS spectra of a system, a standard
ground-state DFT calculation has to be performed first, yielding the
KS orbitals, $\mathrm{u}^\circ_{n,\mathbf{k}}(\mathbf{r})$, energies,
$\varepsilon^\circ_{n,\mathbf{k}}$, for all occupied levels, and the
ground-state charge-density, $\mathrm{n}^\circ(\mathbf{r})$. The information
thus obtained is then used as input for the linear-response
calculation with the \texttt{turboEELS} code. This ground-state
calculation is performed by the \texttt{pw.x} code, which is one of
the key components of the \QE\, package. In
\ref{sec:Sample_input_files} a sample input file for \texttt{pw.x} is
shown for the case of diamond. After successful completion of the
ground-state calculation, the \texttt{pw.x} code writes the
ground-state KS orbitals, energies, and charge-density to disk, together with
all relevant information about the system, like geometry,
pseudopotentials, etc. This data is used by the \texttt{turboEELS}
code which reads all this data at program start. Therefore, it is not
necessary to redefine the system under study in the input file of
\texttt{turbo\_eels.x}.

\subsection{Linear-response TDDFT calculation}   

The linear-response calculation is done using the
\texttt{turbo\_eels.x} code, which performs the Lanczos recursion of
Sec.~\ref{sec:Lanczos_algorithm} for a given transferred momentum, $\mathbf{Q}$. 
This is by far the most time consuming step of the
calculation. In \ref{sec:Sample_input_files} a sample input file for
\texttt{turbo\_eels.x} is shown for the case of diamond. A list of all
input variables of \texttt{turbo\_eels.x} is given in
Table~\ref{tab:Table_input_turbo_eels.x} of
\ref{sec:Input_variables}. The integer input variable \texttt{itermax}
sets up the number of Lanczos iterations, and so determines the
dimension $M$ of the tridiagonal matrix, $^MT_\mathbf{q}$ (see
Sec.~\ref{sec:Lanczos_algorithm}). In fact, one can check whether the
number of iterations is sufficient to achieve an adequately converged
spectrum only at the post-processing level (see
Sec.~\ref{sec:post-proc}). It is possible to add more iterations to
an existing calculation by restarting the
\texttt{turbo\_eels.x} code, setting the parameter
\texttt{restart=.true.} and increasing \texttt{itermax}. The strings
defined in the input 
variables \texttt{prefix} and \texttt{outdir} identify the system data
on disk and must correspond to files created by the \texttt{pw.x}
code.

The input variables \texttt{q1}, \texttt{q2}, and \texttt{q3} are
the three Cartesian components of the transferred momentum, $\mathbf{Q}$,
specified in units of $2\pi/a$, where $a$ is the lattice parameter
specified in the ground-state calculation by \texttt{pw.x}. For each
$\mathbf{Q}$ a separate Lanczos recursion is needed.

The calculation of the charge-density susceptibility, $\chi(\mathbf{Q},
\mathbf{Q}; \omega)$, which is used to obtain the loss function,
$-\mathrm{Im}[\epsilon^{-1}(\mathbf{Q},\omega)]$, can be
performed at different levels of theory, which are specified in the
input using the parameter \texttt{approximation}. This parameter can
take the following values:
$\emph{i)}$ ``\texttt{TDDFT}'' - Time-Dependent Local Density
Approximation (TDLDA) or Time-Dependent Generalized Gradient
Approximation (TDGGA), depending on the XC functional; $\emph{ii)}$ ``\texttt{IPA}'' - the
independent particle approximation, which implies neglecting the
interaction superoperator, $\hat{\mathcal{K}}_\mathbf{q}$, in
Eq.~\eqref{eq:Liouvillian_SBR} (no Hartree and XC terms); and
$\emph{iii)}$ ``\texttt{RPA\_with\_CLFE}'' - Random Phase Approximation (RPA) with
Crystal Local Field Effects (CLFE) (describing an inhomogeneity of
the system), which implies neglecting the response XC potential and leaving
only the response Hartree potential. 

One can choose which flavor of the Lanczos algorithm to use (see
Sec.~\ref{sec:Lanczos_algorithm}). By setting
\texttt{pseudo\_hermitian=.true.}, the pseudo-Hermitian Lanczos
algorithm will be used, otherwise the non-Hermitian Lanczos
biorthogonalization algorithm will be used. It is recommended to use
the former, because it is two times faster.

During the execution of the \texttt{turbo\_eels.x} code, a file named \break
\texttt{prefix.beta\_gamma\_z.dat} will be written to the
\texttt{outdir} directory.
This file contains the Lanczos
coefficients $\beta_i$ and $\gamma_i$, and $z_i$ coefficients needed for the post-processing
calculation. One can use this information for the analysis of the
behavior of these coefficients.

\subsection{Post-processing spectrum calculation}
\label{sec:post-proc}

Once the tridiagonal matrix, $^MT_\mathbf{q}$, is constructed
from the Lanczos coefficients, one can compute the charge-density
susceptibility according to Eqs.~\eqref{eq:post_processing} and
\eqref{eq:resolvent_Liouvillian_Lanczos_2_for_chi}. This task is
performed by the \texttt{turbo\_spectrum.x} program as a
post-processing step, which requires negligible amount of the CPU time
with respect to \texttt{turbo\_eels.x}. This is
so because solving a linear matrix equation~\eqref{eq:post_processing}
and computing a scalar
product~\eqref{eq:resolvent_Liouvillian_Lanczos_2_for_chi} are two
fast operations, which can be performed efficiently using
BLAS and LAPACK libraries.

The code \texttt{turbo\_spectrum.x} is used also for the calculation
of the absorption spectra computed with
\texttt{turboTDDFT}. In order to distinguish the different
applications, it is necessary to set 
\texttt{eels=.true.} in the input for \texttt{turbo\_spectrum.x}.
The labels \texttt{prefix} and \texttt{outdir} identify the system
on disk and must correspond to files created by the
\texttt{turbo\_eels.x} code.

In \ref{sec:Sample_input_files} a sample input file for
\texttt{turbo\_spectrum.x} is shown for the case of diamond. A list of
input variables for the \texttt{turbo\_spectrum.x} program is given in
Table~\ref{tab:Table_input_turbo_spectrum.x} of
\ref{sec:Input_variables}.

In practice, when solving Eq.~\eqref{eq:post_processing}, a small
imaginary part $\eta$ is added to the frequency
argument, $\omega \rightarrow \omega + i\eta$, so as to
regularize the charge-density susceptibility
$\chi(\mathbf{Q},\mathbf{Q};\omega)$
\cite{Rocca:2008, Baroni:2012b}. Setting $\eta$ to a
non-zero value amounts to broadening each individual spectral line or,
alternatively, to convoluting the charge-density susceptibility with a
Lorentzian. The parameter $\eta$ is defined with the
keyword \texttt{epsil}. The EEL/IXS spectra can be computed in any
energy range specified by the keywords \texttt{start} and
\texttt{end}, with a step of energy given by the \texttt{increment}
parameter, all this being specified in Rydberg
(\texttt{units}=0) or electronvolt (\texttt{units}=1) units.

The convergence of the spectrum in the desired energy range can be checked
by varying the number of Lanczos coefficients used.
This number is set by the input keywords \texttt{itermax0} and
\texttt{itermax}. If no extrapolation of Lanczos coefficients is used
(\texttt{extrapolation='no'}), then \texttt{itermax}=\texttt{itermax0}.
These variables can take values up to the number of iterations which
have been calculated with the \texttt{turbo\_eels.x} code.
For a given number of Lanczos iterations, it is possible to improve
the convergence of the computed spectra by extrapolating the
coefficients \cite{Malcioglu:2011}. Such an extrapolation can either be
bi-constant (\texttt{extrapolation='osc'}) or constant
(\texttt{extrapolation='constant'}) \cite{Malcioglu:2011}. In this
case, the input variable \texttt{itermax0} indicates the number of
exact coefficients to be read from file, while \texttt{itermax} is set
to a value which can be chosen arbitrarily large without any
significant computational cost. Such an extrapolation procedure
amounts to increasing the dimension of the tridiagonal matrix,
$^MT_\mathbf{q}$. 

The \texttt{turbo\_spectrum.x} program generates two files, namely, \break
\texttt{prefix.plot\_chi.dat} which contains real and imaginary parts
of the charge-density susceptibility, $\chi(\mathbf{Q}, \mathbf{Q};\omega)$, for each value of the frequency $\omega$, and \break
\texttt{prefix.plot\_eps.dat} which contains real and imaginary parts of the dielectric function and its inverse (see Sec.~\ref{sec:Statement_of_the_problem}). 

The \texttt{turbo\_spectrum.x} program will check the $f$-sum rule according to
Eq.~\eqref{eq:f-sum_definition}. However, in order to obtain a
meaningful result, a convergence with respect to the frequency range
of integration must be checked by the \texttt{start} and \texttt{end}
parameters (the former being equal to zero, and the later being
increased systematically until the convergence). In the
LL approach to TDDFpT, the $f$-sum rule is satisfied
exactly for any number of Lanczos iterations when local
pseudopotentials are used \cite{Baroni:2012b}. However, a violation of
the $f$-sum rule is present when non-local pseudopotentials are used.


\section{Installation instructions and parallelization of the code}
\label{sec:installation_and_parallelization}

\subsection{Installation instructions}
\label{sec:Installation_instructions}

The \texttt{turboEELS} program is distributed as source code, like the
other components of the \QE\, distribution. The installation procedure
is the same for all modules in the \QE\,
package. \QE\, and \texttt{turboEELS} make use of GNU autotools. The
\texttt{TDDFPT} repository, which contains the source
\texttt{turboEELS} code (and \texttt{turboTDDFT} code), must be
residing within the \QE\, tree. The code is compiled with the
following commands from within the \QE\,tree:
\begin{equation*}
 \begin{array}{c}
   \texttt{./configure} \\ \texttt{make pw} \qquad\quad
   \\ \texttt{make tddfpt}
 \end{array}
\end{equation*}
Here, the first step sets up the environment (compilers, libraries,
etc.). The second step compiles the \texttt{pw.x} code and creates a
link to this executable in the \texttt{bin/} repository of the \QE\,
tree. In the third step, the \texttt{turboEELS} codes
(\texttt{turbo\_eels.x} and \texttt{turbo\_spectrum.x}) are compiled,
together with the \texttt{turboTDDFT} codes (\texttt{turbo\_lanczos.x} and
\texttt{turbo\_davidson.x}). Links to all these executables are created
in the \texttt{bin/} directory of the \QE\, tree. 
Further detailed installation instructions can be found in the
documentation that comes with the \QE\, distribution. 

The \texttt{turboEELS} code is tightly bound to the \texttt{pw.x} code
(residing in \texttt{PW}), to the \texttt{ph.x} code (residing in
\texttt{PHonon}), and to the \texttt{turboTDDFT} code.

\subsection{Parallelization}

Like the other components of the \QE\, package, the \texttt{turboEELS}
code is optimized to run on a variety of different platforms, from laptops
to massively parallel architectures. The parallelization of the
\texttt{turboEELS} code is achieved by using the message-passing paradigm
and calls to standard Message Passing Interface (MPI) libraries
\cite{MPI:1994}. High performance on massively parallel architectures
is achieved by distributing both data and computations in a
hierarchical way across processors. The \texttt{turboEELS} code
supports two levels of parallelization: $\emph{i)}$ a plane-wave
parallelization, which is implemented by distributing real- and
reciprocal-space grids across the processors, and 
$\emph{ii)}$~a~$\mathbf{k}$~points parallelization, which is implemented by dividing
all processors into pools, each taking care of one or more
$\mathbf{k}$ points. The Fast Fourier Transforms (FFT's), which are
used for transformations from real space to reciprocal space and vice
versa, are also efficiently parallelized among processors.


\section{Benchmarking}
\label{sec:benchmarking}

We now proceed to the validation of the \texttt{turboEELS} code by
calculating the loss function in bulk diamond. We will make a
comparison with Ref.~\cite{Waidmann:2000}, where a
theoretical study (using the conventional TDDFT approach) is
confronted with the experimental EEL spectra. 
Our purpose here is not to analyze the EEL
spectra of diamond or to achieve a remarkable agreement with
experiment, but to demonstrate the correctness of our implementation
of \texttt{turboEELS} by comparing the computed EEL spectrum with the
one obtained using a conventional TDDFT approach, and to show the
convergence behavior of the EEL spectra when using the LL
approach to EEL/IXS.

\begin{figure*}[t]
  \begin{center}
    \subfigure[]{\includegraphics[width=0.47\textwidth]{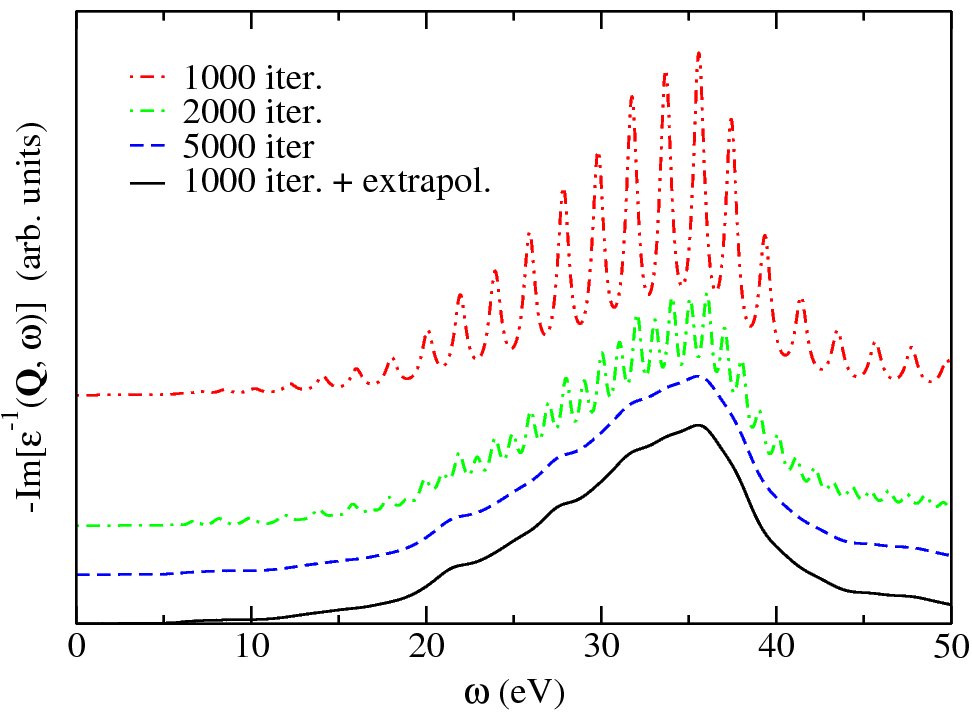}
      \label{fig:diamond_conv_iter}}
      \hspace{0.09 cm}
      \subfigure[]{\includegraphics[width=0.47\textwidth]{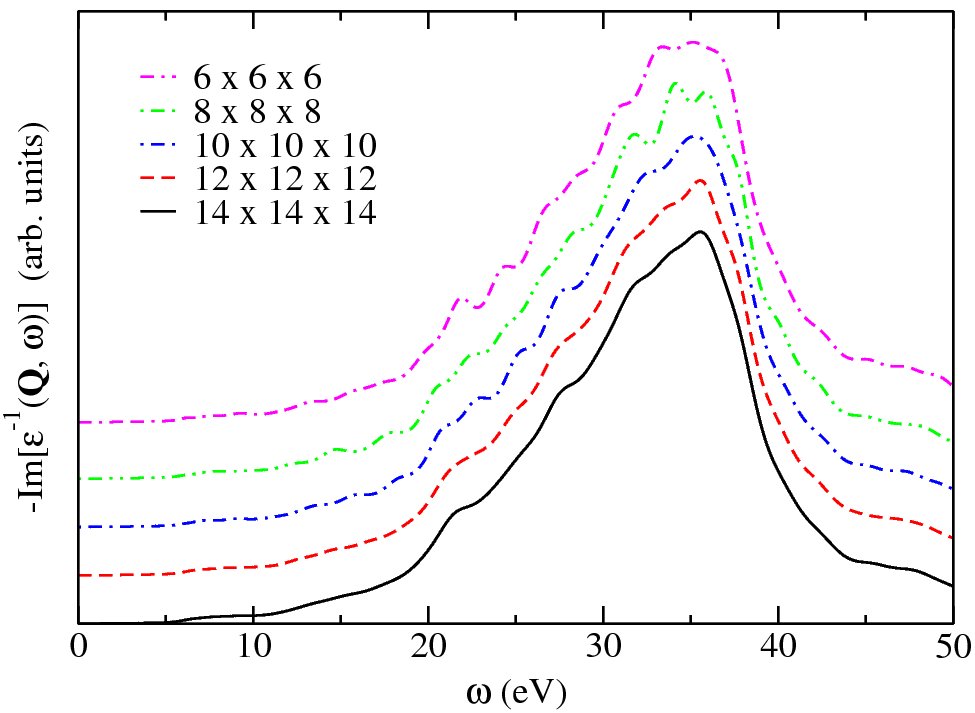}
      \label{fig:diamond_conv_k}}
    \caption{Loss function of diamond at $Q=0.15$~\AA$^{-1}$ along the
      [100] direction.  (a)~Convergence with respect to the number of
      Lanczos iterations, and effect of the extrapolation technique,
      using a $14 \times 14 \times 14$ Monkhorst-Pack $\mathbf{k}$
      point mesh. (b) Convergence with respect to the size of the
      $\mathbf{k}$ point mesh, for 1000 Lanczos iterations using
      extrapolation technique.  Curves have been shifted vertically
      for clarity.}
    \label{fig:convergence}
  \end{center}
\end{figure*}

We have chosen the parameters of our calculations as close as possible to
those of Ref.~\cite{Waidmann:2000} in order to simplify the comparison.
We have used the experimental lattice parameter of 3.57~\AA\,
\cite{Wyckoff:1963}, and used the local density approximation
(LDA) with the Perdew-Zunger parametrization of the electron-gas data
\cite{Perdew:1981}. A norm-conserving pseudopotential was employed from
the \QE\, database \cite{Timrov:Note:2014:PP}, and we have used a
kinetic-energy cutoff of 50~Ry. The first BZ has been sampled with a
$14 \times 14 \times 14$ Monkhorst-Pack (MP) $\mathbf{k}$-point mesh
\cite{Monkhorst:1976}, resulting in 280 $\mathbf{k}$-points in the
irreducible wedge of BZ for the ground-state calculation, and 2940
$\mathbf{k}$-points in the irreducible wedge of BZ for the small group
of $\mathbf{Q}$ for the linear-response TDDFT calculation. We have
used a Lorentzian broadening of $\eta = 0.03$~Ry for the
charge-density susceptibility (loss function). All these parameters must be specified
in the following way as an input for the calculation, see
\ref{sec:Sample_input_files}. All the results were obtained at the TDLDA
level, {\it i.e.} inlcuding CLFE and XC effects.

Figure~\ref{fig:convergence} shows the convergence of the loss function
of diamond with respect to the number of Lanczos iterations and
$\mathbf{k}$ points in the BZ, for $Q=0.15$~\AA$^{-1}$ along the [100]
direction. In Fig.~\ref{fig:diamond_conv_iter} it can be seen that
after 1000 Lanczos iterations the loss function shows spurious
wiggles, which disappear by increasing the number of Lanczos
iterations up to 5000. However, the speed of the convergence can be
increased by using the extrapolation technique \cite{Rocca:2008,
  Malcioglu:2011}. Indeed, convergence can be reached after 1000
Lanczos iterations when the extrapolation of Lanczos coefficients is
used (up to $20000$ in this case). In Fig.~\ref{fig:diamond_conv_k}
one can see that the convergence of the loss function with respect to
the number of $\mathbf{k}$ points (size of the MP mesh) is reached for
the $14 \times 14 \times 14$ MP mesh.

\begin{figure*}[t]
  \begin{center}
    \subfigure[]{\includegraphics[width=0.48\textwidth]{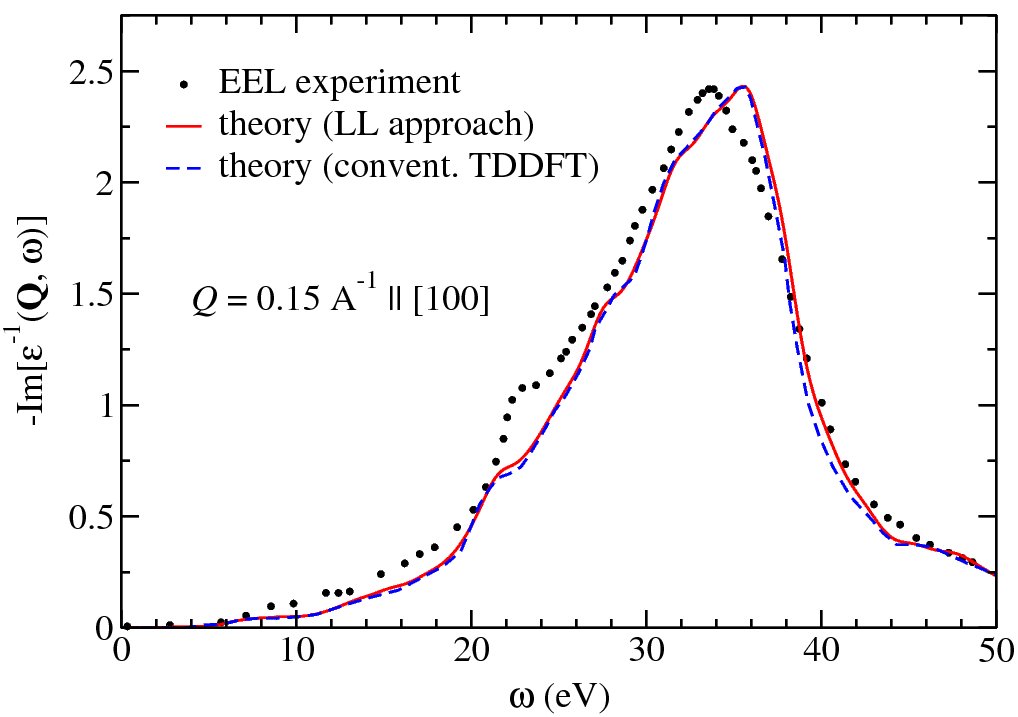}
      \label{fig:diamond_theor_vs_exp}}
      \hspace{0.1 cm}
      \subfigure[]{\includegraphics[width=0.4625\textwidth]{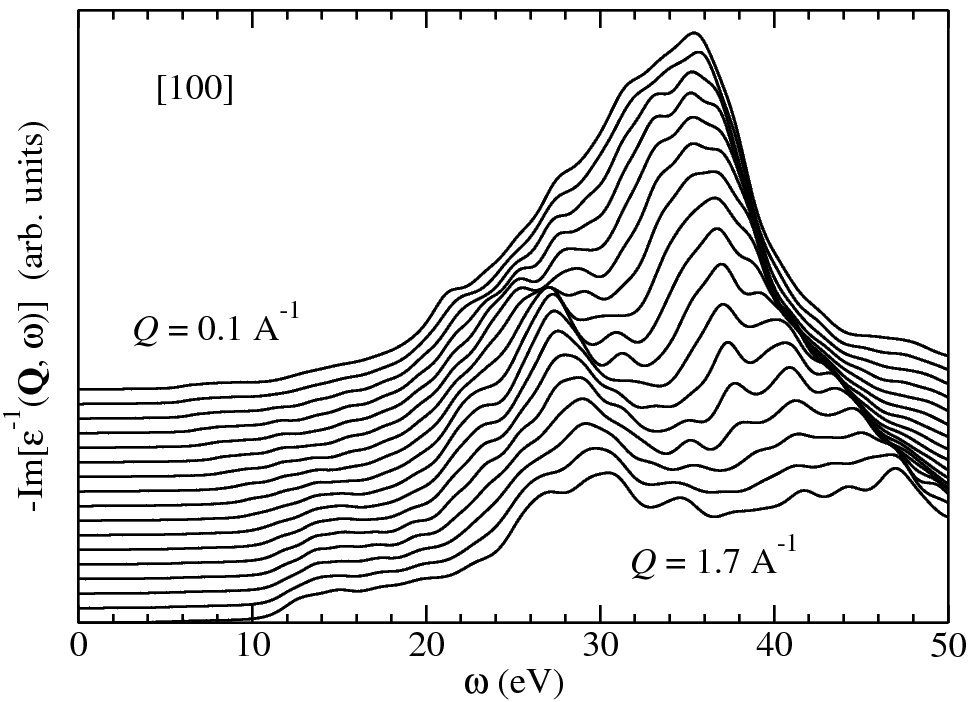}
      \label{fig:diamond_dispersion}}
    \caption{(a) Comparison of the loss function of diamond for
      $Q=0.15$~\AA$^{-1}$ along the [100] direction
      calculated using the Liouville-Lanczos (LL) approach, with
      experiment and with previous calculations \cite{Waidmann:2000}.
      (b) Loss function calculated using the LL approach for various
      transferred momenta $Q$ along [100] ranging from
      0.1~\AA$^{-1}$ to 1.7~\AA$^{-1}$ with a step 0.1~\AA$^{-1}$.}
     \label{fig:comparison_with_theor_and_exp}
   \end{center}
\end{figure*}

In Fig.~\ref{fig:comparison_with_theor_and_exp} we compare our result
obtained using the LL approach to EEL/IXS with those
obtained using the conventional TDDFT approach based on the solution
of the Dyson-like equation and with experiment \cite{Waidmann:2000},
for $Q=0.15$~\AA$^{-1}$ along the [100] direction. The
agreement between the LL approach and previous theoretical
calculations is excellent, which validates the LL approach. It has the
same level of accuracy as the conventional TDDFT approach, but
requires much smaller computational cost. The agreement between theory
and experiment is remarkable, though there are some small
deviations. In Fig.~\ref{fig:diamond_theor_vs_exp}, the plasmon peak
is around at 35~eV, the peak due to interband transitions is around
22~eV. The shoulders at 28~eV and 32~eV are not seen in the experimental
spectrum due to the large broadening, but are resolved in the
theoretical spectrum. 
Figure~\ref{fig:diamond_dispersion} shows the loss function of diamond
for various transferred momenta $Q$ along the [100] direction
ranging from 0.1~\AA$^{-1}$ to 1.7~\AA$^{-1}$ with a step
0.1~\AA$^{-1}$. One can see a strong dispersion of the plasmon peak
and its damping due to entrance in the electron-hole continuum, a weak
dispersion of the peak at 22~eV due to interband transitions and its
fast damping, and an appearance of the new peak at
$Q=0.8$~\AA$^{-1}$ around at 26~eV which also shows a weak
dispersion. A detailed analysis of the loss function of diamond and its dispersion along various 
directions can be found in Ref.~\cite{Waidmann:2000}.

\begin{figure}[h!]
  \begin{center}
    \includegraphics[width=0.48\textwidth]{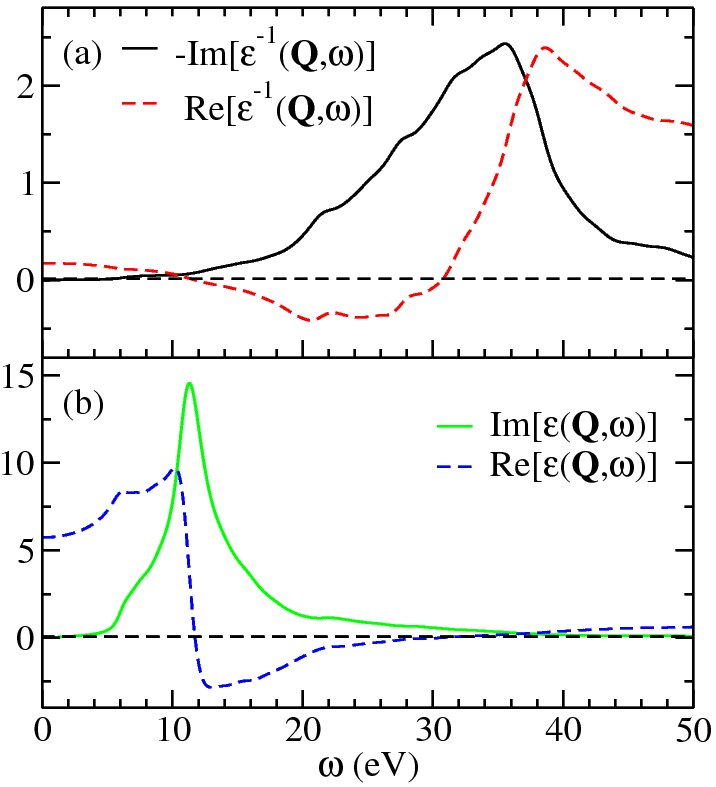}
    \caption{Real and imaginary parts of the inverse dielectric function (a) and the dielectric function (b) for 
      $Q=0.15$~\AA$^{-1}$ along the [100] direction
      calculated using the Liouville-Lanczos (LL) approach.}
     \label{fig:comparison_epsm1_and_eps}
   \end{center}
\end{figure}

From the knowledge of the real and imaginary parts of the inverse dielectric function for a given transferred momentum, $\epsilon^{-1}(\mathbf{Q},\omega)$, we have computed the real and imaginary parts of the dielectric function, $\epsilon(\mathbf{Q},\omega)$, as explained in Sec.~\ref{sec:Statement_of_the_problem}. The result is illustrated in Fig.~\ref{fig:comparison_epsm1_and_eps}. As can be seen, the plasmon at $\sim 35$~eV occurs when $\mathrm{Re}[\epsilon(\mathbf{Q},\omega)]$ is almost zero and changes its sign with a positive slope, and $\mathrm{Im}[\epsilon(\mathbf{Q},\omega)]$ is very small. The strongest peak in $\mathrm{Im}[\epsilon(\mathbf{Q},\omega)]$ occurs at $\sim 11$~eV, which is due to interband transitions, and occuring when $\mathrm{Re}[\epsilon(\mathbf{Q},\omega)]=0$ and changes its sign with a negative slope. And the peak in the loss function $-\mathrm{Im}[\epsilon^{-1}(\mathbf{Q},\omega)]$ at $\sim 22$~eV is indeed due to interband transitions, because there is also a weak peak in $\mathrm{Im}[\epsilon(\mathbf{Q},\omega)]$ around the same frequency. A detailed discussion about the origin of the peaks in the EEL spectrum of diamond can be found in Ref.~\cite{Waidmann:2000}. 

Finally, we have checked the $f$-sum rule according to Eq.~\eqref{eq:f-sum_definition}. We have found that it is satisfied with an extremely small violation of $<1\%$, which is due to a non-locality of the pseudopotential used in our calculations \cite{Baroni:2012b}.

\section{Conclusions}
\label{sec:Conclusions}

We have presented the \texttt{turboEELS} code that implements a
Liouville-Lanczos approach to time-dependent density functional
perturbation theory for the computation of EEL and IXS spectra for any
finite transferred momentum $\mathbf{Q}$. We have presented a
theoretical description of the approach and its implementation as part
of the \QE\, package.

The \texttt{turboEELS} code has a numerical scalability of the same
order as standard ground-state DFT calculations. It does
not require the calculation of empty electronic states, and a
computationally expensive inversion of the dielectric matrix is
replaced by a very efficient recursive Lanczos algorithm, which allows
us to compute EEL/IXS spectra in a wide energy range. These advances
allow us to compute EEL/IXS spectra of complex systems, {\it e.g.}
high-Miller-index surfaces. 

In the same spirit as the \QE\, project, \texttt{turboEELS} provides
scientists worldwide with a well commented and open-source framework for
implementing their ideas. It is in our best hope that
\texttt{turboEELS} can benefit from the already well established users
community of \QE\, for incorporating new ideas and keep growing in the
future. The \texttt{turboEELS} code is hosted in a community
accessible SVN repository \cite{QEFORGE} and hence, apart from
releases in \QE, those who are willing to test the latest experimental
implementations are welcome to do so and to contribute with their
feedback.

The \texttt{turboEELS} code can be extended so as to employ hybrid XC
functionals (which are already supported by the \texttt{turboTDDFT}
code at $\mathbf{k}=0$ \cite{Ge:2014}), which would allow us to
describe excitons and their dispersion \cite{Gatti:2013}.

\section*{Acknowledgements}

We thank S.~de~Gironcoli, A.~Dal~Corso, M.~Saitta, L.~Reining, and T.~Gorni for
valuable discussions. The work of I.T. and N.V. has been performed
under the auspices of the \emph{Laboratoire d'excellence en
  nanosciences et nanotechnologies Labex Nanosaclay}. Support from the
ANR (Project PNANO ACCATTONE) and from DGA are gratefully
acknowledged. Computer resources were provided by GENCI (Project
No. 2210) and CINECA (Project ISCRA~C ``t-EELS\_E''). Technical
support from C.~Cavazzoni and F.~Affinito at CINECA is gratefully
acknowledged.

\appendix

\section{Input Variables}
\label{sec:Input_variables}


\begin{table}[H]
 \begin{tabular}{|>{\centering}m{0.6cm}|c}
      \cline{1-1} 
      Card  & 
      \begin{tabular}{|>{\centering}m{3.45cm}|>{\centering}m{1.5cm}|>{\centering}p{9.0cm}|} \hline 
      Variable name  & Default Value  & Description \tabularnewline
      \end{tabular} \tabularnewline
      \cline{1-1} 
\begin{sideways}
   \textbf{lr\_input} 
\end{sideways}
      & \begin{tabular}{|>{\raggedright}m{3.45cm}|> {\centering}m{1.5cm}|m{9.0cm}|} \hline 
        \texttt{prefix} & '\texttt{pwscf}' & {\footnotesize 
          The files generated by the ground state \texttt{pw.x} run should have this same prefix.}\tabularnewline 
        \texttt{outdir} & './' & {\footnotesize
          Working directory. On start, it should contain the files generated by a ground state \texttt{pw.x} run.}\tabularnewline 
        \texttt{restart} & \emph{.false.} & {\footnotesize 
          When set to \emph{.true.}, \texttt{turbo\_eels.x} will attempt to restart from a previous interrupted calculation.  
          (see \texttt{restart\_step} variable).} \tabularnewline
        \texttt{restart\_step} & \texttt{itermax} & {\footnotesize 
          The code writes restart files every \texttt{restart\_step} iterations. Restart files are automatically written at the end of
          \texttt{itermax} Lanczos steps.} \tabularnewline 
        \texttt{lr\_verbosity} & 1 & {\footnotesize Verbosity level} \tabularnewline 
      \end{tabular} \tabularnewline
      \cline{1-1} 
\begin{sideways}
  \textbf{lr\_control}  
\end{sideways}
      & \begin{tabular}{|>{\raggedright}m{3.45cm}|>{\centering}m{1.5cm}|m{9.0cm}|} \hline 
        \texttt{itermax} & 500 & {\footnotesize Number of iterations to be performed.} \tabularnewline 
        \texttt{q1, q2, q3}    & 1, 1, 1 & {\footnotesize Cartesian components of the transferred momentum in units of 
          $2 \pi/a$ (where $a$ is the lattice parameter of the unit cell).} \tabularnewline 
        \texttt{approximation} & '\texttt{TDDFT}'  & {\footnotesize A string describing an approximation:
                               '\texttt{TDDFT}' - Time-Dependent Local Density Approximation or 
                                                  Time-Dependent Generalized Gradient Approximation 
                                                  (depending on the XC functional used),
                               '\texttt{IPA}' - Independent Particle Approximation,
                               '\texttt{RPA\_with\_CLFE}' - Random Phase Approximation with Crystal Local Field Effects.}
                                \tabularnewline
        \texttt{pseudo\_hermitian} & \emph{.true.} & \footnotesize{If \emph{.true.} then the pseudo-Hermitian Lanczos algorithm is used, if \emph{.false.} then the non-Hermitian Lanczos biorthogonalization algorithm is used (which is two times slower). 
          } \tabularnewline \hline
      \end{tabular} \tabularnewline
      \cline{1-1}
 \end{tabular} 
 \caption{Input variables for \texttt{turbo\_eels.x} }
 \label{tab:Table_input_turbo_eels.x}
\end{table}


\begin{table}[H]
 \begin{tabular}{|>{\centering}m{0.6cm}|c}
      \cline{1-1}
      Card  &
      \begin{tabular}{|>{\centering}m{3.45cm}|>{\centering}m{1.5cm}|>{\centering}p{9.0cm}|} \hline
      Variable name  & Default Value  & Description \tabularnewline
      \end{tabular} \tabularnewline
      \cline{1-1}
\begin{sideways}
   \textbf{lr\_input}
\end{sideways}
      & \begin{tabular}{|>{\raggedright}m{3.45cm}|> {\centering}m{1.5cm}|m{9.0cm}|} \hline
        \texttt{prefix} & '\texttt{pwscf}' & {\footnotesize
          Prefix of the files generated by the previous \texttt{turbo\_eels.x} run.}\tabularnewline
        \texttt{outdir} & './' & {\footnotesize
          The directory where the output files produced by the previous \texttt{turbo\_eels.x} run are stored.}\tabularnewline
        \texttt{eels} & \emph{.false.} & {\footnotesize
          Must be set to \emph{.true.} for EELS. EELS-specific operations will be performed.}\tabularnewline
        \texttt{itermax0} & 1000 & {\footnotesize
          Number of Lanczos coefficients to be read from the file.}\tabularnewline
        \texttt{itermax} & 1000 & {\footnotesize
          The total number of Lanczos coefficients that will be considered in the calculation of the charge-density 
          susceptibility (loss function). If \texttt{itermax} $>$ \texttt{itermax0}, the Lanczos coefficients in between 
          \texttt{itermax0}+1 and \texttt{itermax} will be extrapolated.}\tabularnewline
        \texttt{extrapolation} & '\texttt{no}' & {\footnotesize
          Sets the extrapolation scheme for Lanczos coefficients. '\texttt{osc}' = bi-constant extrapolation; 
          '\texttt{constant}' = constant extrapolation; '\texttt{no}' = no extrapolation.}\tabularnewline
        \texttt{epsil} & 0.02 & {\footnotesize
          The Lorentzian broadening parameter (in Rydberg units).}\tabularnewline
        \texttt{units} & 0 & {\footnotesize
          Unit system used. 0: Rydbergs; 1: Electronvolts}\tabularnewline
        \texttt{start} & 0.0 & {\footnotesize
          The charge-density susceptibility and the loss function are computed starting from this value.
          In units set by the \texttt{units} variable.}\tabularnewline
        \texttt{end} & 2.5 & {\footnotesize
          The charge-density susceptibility and the loss function are computed up to this value.
          In units set by the \texttt{units} variable.}\tabularnewline
        \texttt{increment} & 0.001 & {\footnotesize
          Incremental step used to define the mesh between \texttt{start} and \texttt{end}. 
          In units set by the \texttt{units} variable.}\tabularnewline
        \texttt{verbosity} & 0 & {\footnotesize 
          This integer variable controls the output verbosity.} \tabularnewline \hline
      \end{tabular} \tabularnewline
      \cline{1-1}
 \end{tabular}
 \caption{Input variables for \texttt{turbo\_spectrum.x} }
 \label{tab:Table_input_turbo_spectrum.x}
\end{table}


\section{Sample input files}
\label{sec:Sample_input_files}

\vskip 0.5 cm

\noindent
{\bf Input example 1:} Input sample for \texttt{pw.x}
\begin{verbatim}
&control
   calculation = 'scf'
   restart_mode='from_scratch',
   pseudo_dir = './pseudo',
   outdir='./out',
   prefix='diamond'
/
&system
   ibrav = 2,
   celldm(1) = 6.75,
   nat = 2,
   ntyp = 1,
   ecutwfc = 50.0
/
&electrons
   diagonalization='david'
   mixing_mode = 'plain'
   mixing_beta = 0.7
   conv_thr =  1.0d-12
/
ATOMIC_SPECIES
 C  12.011  C.pz-vbc.UPF
ATOMIC_POSITIONS {alat}
 C 0.00 0.00 0.00
 C 0.25 0.25 0.25
K_POINTS {automatic}
 14 14 14 1 1 1
\end{verbatim}

\vskip 0.5 cm

\noindent
{\bf Input example 2:} Input sample for \texttt{turbo\_eels.x}
\begin{verbatim}
&lr_input
   prefix='diamond',
   outdir='./out',
   restart_step = 250,
   restart = .false.
/
&lr_control
   itermax = 500,
   q1 = 0.085,
   q2 = 0.000,
   q3 = 0.000,
/
\end{verbatim}

\vskip 0.5 cm

\noindent
{\bf Input example 3:} Input sample for \texttt{turbo\_spectrum.x}
\begin{verbatim}
&lr_input
   prefix='diamond',
   outdir='./out',
   eels = .true.
   itermax0 = 500
   itermax  = 20000
   extrapolation = "osc"
   epsil = 0.03
   units = 1
   start = 0.0d0
   increment = 0.01d0
   end = 50.0d0
   verbosity = 0
/
\end{verbatim}


\end{document}